\documentclass[10pt, twocolumn, journal]{IEEEtran}

\usepackage{epsf}
\usepackage{graphics}
\usepackage{times}
\usepackage{graphicx}
\usepackage{amsmath}
\usepackage{amssymb}
\usepackage{epsfig}
\usepackage{subfigure}
\usepackage{array}
\usepackage{float}
\usepackage{bbm}

\begin{document}

\title{On the Shift Value Set of Cyclic Shifted Sequences for PAPR Reduction in OFDM Systems}

\author{Kee-Hoon Kim
\thanks{The author is currently with Samsung Electronics, Co., Ltd., Suwon, Gyeonggi-do, Korea.}
\thanks{email: kkh@ccl.snu.ac.kr}
}

\maketitle
\begin{abstract}
Orthogonal frequency division multiplexing (OFDM) signals have high peak-to-average power ratio (PAPR), which causes distortion when OFDM signal passes through a nonlinear high power amplifier (HPA). A partial transmit sequence (PTS) scheme is one of the typical PAPR reduction methods. A cyclic shifted sequences (CSS) scheme is evolved from the PTS scheme to improve the PAPR reduction performance, where OFDM signal subsequences are cyclically shifted and combined to generate alternative OFDM signal sequences. The shift value (SV) sets in the CSS scheme should be carefully selected because those are closely related to the PAPR reduction performance of the CSS scheme. In this letter, we propose some criteria to select the good SV sets and verify its validness through simulations.
\end{abstract}


\section{Introduction}
Orthogonal frequency division multiplexing (OFDM) is a multicarrier modulation method
utilizing the orthogonality of subcarriers. OFDM has been adopted as a standard modulation
method in many wireless communication systems such as digital audio broadcasting (DAB),
digital video broadcasting (DVB), IEEE 802.11 wireless local area network (WLAN), and IEEE
802.16 wireless metropolitan area network (WMAN). Similar to other multicarrier schemes,
OFDM has a high peak-to-average power ratio (PAPR) problem, which makes its implementation
quite costly. Thus, it is highly desirable to reduce the PAPR of OFDM signal sequences.
Over the last few decades, various schemes to reduce the PAPR of OFDM signal sequences
have been proposed such as clipping, coding, active constellation extension
(ACE) \cite{Krongold}, tone reservation (TR), partial transmit sequence (PTS) \cite{Muller}, and selected
mapping (SLM) \cite{Badran}--\cite{Jiang}.

Like the SLM scheme, the PTS scheme statistically improves the characteristic of the PAPR distribution of OFDM signals without signal distortion. In the PTS scheme, the input symbol sequence is partitioned into a number of disjoint input symbol subsequences. Inverse fast Fourier transform (IFFT) is then applied to each input symbol subsequence and the resulting OFDM signal subsequences are combined after being multiplied by a set of rotation factors. Next the PAPR is computed for each resulting sequence and then the OFDM signal sequence with the minimum PAPR is transmitted.

Hill et al. proposed a cyclic shifted sequences (CSS) scheme, where cyclic shift is used instead of multiplying rotation factor to the OFDM signal subsequences \cite{Hill},\cite{Hill2}, and the CSS scheme can be viewed as a special case of the PTS scheme \cite{Lu}. It is widely known that the CSS scheme is better than the PTS scheme from every aspect. First, its PAPR reduction performance is better than the PTS scheme's. Second, it is possible to recover the transmitted OFDM signal sequence without side information using some additional techniques at the receiver \cite{Yang},\cite{Long}.

In this letter, we investigate how to select the shift value (SV) sets in order to boost the PAPR reduction performance of the CSS scheme. We introduce some criteria to select the good SV sets considering the autocorrelation function (ACF) of OFDM signal subsequences, and then verify its validness through simulations.

\section{OFDM System and PAPR}
In an OFDM system, an OFDM signal sequence in time domain is generated by IFFT as
\begin{equation}
x(n) = \frac{1}{\sqrt{N}}\sum_{k=0}^{N-1} X(k) e^{j \frac{2\pi k n}{N}}
\end{equation}
where $N$ is the number of subcarriers, $X = \{X(0),X(1),\cdots,X(N-1)\}$ is an input symbol sequence in frequency domain, and $x = \{x(0),x(1),\cdots,x(N-1)\}$ is an OFDM signal sequence in time domain.
The PAPR of the OFDM signal sequence $x$ is defined as
\begin{equation}
\mathrm{PAPR} = \frac{\max_{0\leq n <N} |x(n)|^2}{E\{|x(n)|^2\}}
\end{equation}
where $E\{\cdot\}$ represents the expectation.

\section{Cyclic Shifted Sequences (CSS)}
\begin{figure}[h]
\centering
\includegraphics[width=\linewidth]{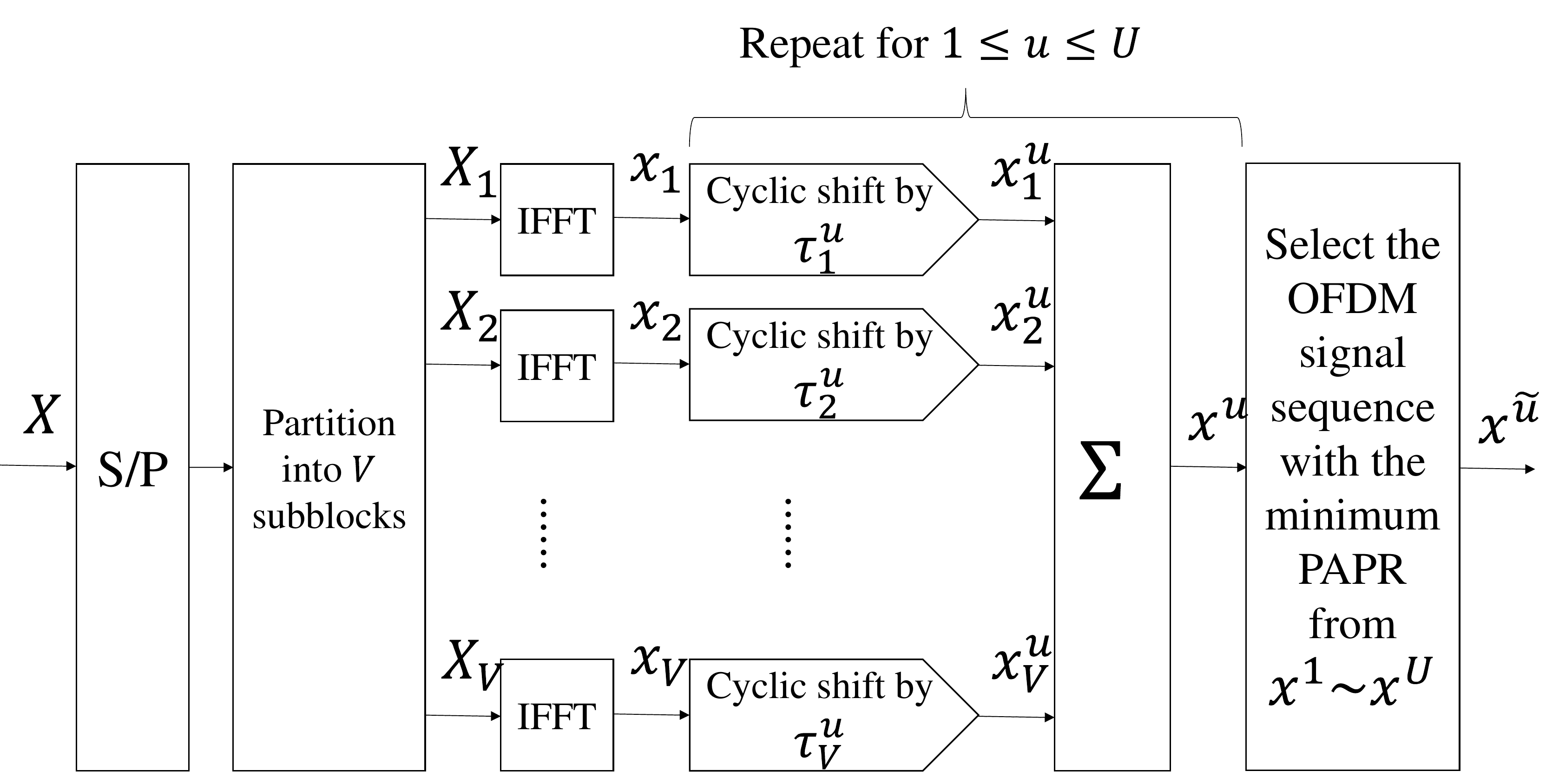}
\caption{A block diagram of the CCS scheme \cite{Hill}.}
\label{fig:CCSblock}
\end{figure}

Fig. \ref{fig:CCSblock} shows a block diagram of the CCS scheme testing $U$ alternative OFDM signal sequences in total \cite{Hill}.
In the CCS scheme, $X$ is divided by a certain partitioning pattern into $V$ disjoint subblocks, input symbol subsequences $X_1,X_2,\cdots,X_V$. Then IFFT converts the $V$ subblocks in frequency domain to the $V$ OFDM signal subsequences in time domain $x_1,x_2,\cdots,x_V$, where $x_v = \{x_v(0),x_v(1),\cdots,x_v(N-1)\}$, $1\leq v \leq V$. For simplicity, we assume that both $N$ and $V$ are integers of power of two.
After that, the $V$ OFDM signal subsequences are cyclically shifted and combined together to make the $u$-th ($1\leq u \leq U$) alternative OFDM signal sequence as
\begin{equation}
x^u = \sum_{v=1}^{V} x_v^u
\end{equation}
where $x_v^u$ denotes the leftward cyclically shifted version of $x_v$ by some integer $\tau_v^u$ ($1\leq v \leq V$). As the SLM or PTS schemes, the candidate with the lowest PAPR, $x^{\tilde{u}}$, is chosen by exhaustive search for transmission with $\lceil\log_2 U \rceil$ bits side information. By using some additional techniques at the receiver, the side information can be recovered \cite{Yang}.

The cyclic shift operation does not destroy the orthogonality between the input symbols $X(k)$'s because, as we all know, cyclic shifting in time domain is equivalent to multiplying a corresponding linear phase vector in frequency domain \cite{Hill}.
In this letter, we denote $\tau_v^u$ as \textit{a shift value} and also denote $\overline{\tau}^u=\{\tau_1^u,\tau_2^u,\cdots,\tau_V^u\}$ as \textit{a SV set} for the $u$-th alternative OFDM signal sequence. Clearly, we have to construct $U$ SV sets ($\overline{\tau}^1,\overline{\tau}^2,\cdots,\overline{\tau}^U$) to implement the CCS scheme testing $U$ alternative OFDM signal sequences.

In this letter, we consider three partition methods, which are random, adjacent, and interleaved partition methods. The random partition method gives the best PAPR reduction performance among them while the interleaved partition method gives the worst PAPR reduction performance but it needs the lowest computational complexity.

\section{Desirable Shift Value Sets in the CSS Scheme}
In the CSS scheme, the PAPR reduction performance depends on how to construct $U$ SV sets $\{\overline{\tau}^1,\overline{\tau}^2,\cdots,\overline{\tau}^U$\}. Considering the fact that the true objective of the CCS scheme is to reduce the probability of the PAPR exceeding some
threshold level rather than to reduce the PAPR of each alternative
OFDM signal sequence itself, we may say in general that $U$ SV sets that make alternative OFDM signal sequences as statistically independent as possible can perform well.

There are $N^V$ cases of one SV set. That is, $\overline{\tau}^u = \{\tau_1^u,\tau_2^u,\cdots,\tau_V^u\}$ can be varied from $\{0,0,\cdots,0\}$ to $\{N-1,N-1,\cdots,N-1\}$. Among these $N^V$ possible SV sets, we select only $U$ SV sets in the CSS scheme. In general, $N^V$ is a huge number and thus it is hard to design $U$ SV sets without any criterion, which motivates us to propose criteria to select good $U$ SV sets in this letter.

\subsection{Desirable Shift Value Sets without Consideration of Correlation}
Suppose that the components in the OFDM signal subsequence are mutually independent. That is, $x_v(0),x_v(1),\cdots,x_v(N-1)$ are mutually independent for all $v$. If $U=2$ and $V=4$, we can select two SV sets, $\overline{\tau}^1 = \{0,0,0,0\}$ and $\overline{\tau}^2 = \{0,0,0,1\}$. In this case, the PAPR reduction performance becomes not good because two alternative OFDM signal sequences ($x^1$ and $x^2$) generated by using theses two SV sets may have high dependency each other. Instead, it is better to select two SV sets such as $\overline{\tau}^1 = \{0,0,0,0\}$ and $\overline{\tau}^2 = \{0,1,2,3\}$, which leads to increasing statistically independency between two alternative OFDM signal sequences ($x^1$ and $x^2$). That is, in order to generate two alternative OFDM signal sequence with independency, the relative distances $\tau_v^1-\tau_v^2$ for all $v$'s have to be distinct from each other. When $U >2$, this has to be guaranteed for all possible SV set pairs out of $U$ SV sets. Now we obtain the following criterion.

\textbf{\emph{Criterion 1}} : Suppose that we have $U$ SV sets; For every $(i,j)$ pair out of the $U$ SV sets ($i\neq j$), the pair should satisfy the condition that the relative distances $\tau_v^i - \tau_v^j \mod N$ are distinct from each other for all $v$'s.

Note that the $\emph{Criterion 1}$ is valid when the components in all alternative OFDM signal subsequences are mutually independent.
However, actually the OFDM signal subsequence components are not mutually independent because the corresponding input symbol subsequences in frequency domain have $N-N/V$ zeros. Cyclically shifting the OFDM signal subsequence of which the components are correlated may make the resulting OFDM signal subsequence similar to the original one. For example, suppose that the ACF $R_{x_v}(m)$ of the $v$-th OFDM signal subsequence $x_v$ has a peak value when $m=1$. Then cyclically shifting by one cannot make $x_v$ much different.
Therefore, we have to consider the ACF of $x_v$ additionally.

\subsection{ACF of OFDM Signal Subsequences}
Let $S_v$ be the discrete power spectrum of the $v$-th OFDM signal subsequence $x_v$, namely,
\begin{equation}
S_v = \{p(0),p(1),\cdots,p(N-1)\}
\end{equation}
where
$p(k)\doteq E\{|X_v(k)|^2\}$, and $p(k)$ can have the value of zero or one. This is due to the assumption that the modulation order of all subcarriers is equal and the average power is normalized to one. For example, if the interleaved partition is used, $S_1=\{1 0 1 0 1 0 1 0\}$ and $S_2=\{0 1 0 1 0 1 0 1\}$ when $N=8$ and $V=2$.

Then the ACF $R_{x_v}(m)$ is given by inverse discrete Fourier transform (IDFT) of $S_v$. Considering the input symbol sequence $X_v$ has $N-N/V$ zeros in a certain pattern, the corresponding ACF $R_{x_v}(m)$ has a specific shape. Here we investigate only the magnitude of the ACF because the high peak of the OFDM signal sequence is closely related to the magnitude of components.

\subsubsection{For Interleaved Partition}
In this case, $S_v$ is an impulse train with an interval of $V$. Then, the ACF also becomes the impulse train as
\begin{equation}\label{eq:intACF}
|R_{x_v}(m)| =
\begin{cases}
\frac{\sqrt{N}}{V} &\textrm{if } m = 0\mod \frac{N}{V} \\
0 &\textrm{otherwise}.
\end{cases}
\end{equation}

\subsubsection{For Adjacent Partition}
In this case, $S_v$ is a rectangular function with a width of $N/V$. Then the ACF becomes the function as
\begin{equation}\label{eq:adACF}
|R_{x_v}(m)| =
\begin{cases}
\frac{\sqrt{N}}{V} &\textrm{if } m = 0\\
\frac{\sin(m \pi /V)}{\sqrt{N} \sin(m\pi /N)} &\textrm{if } m \neq 0.
\end{cases}
\end{equation}

\subsubsection{For Random Partition}
In this case, $S_v$ can be viewed as a binary pseudo random sequence. Then the ACF has a shape similar to a delta function, where the components except $m=0$ are close to zero.

\begin{figure}[h]
\centering
\includegraphics[width=\linewidth]{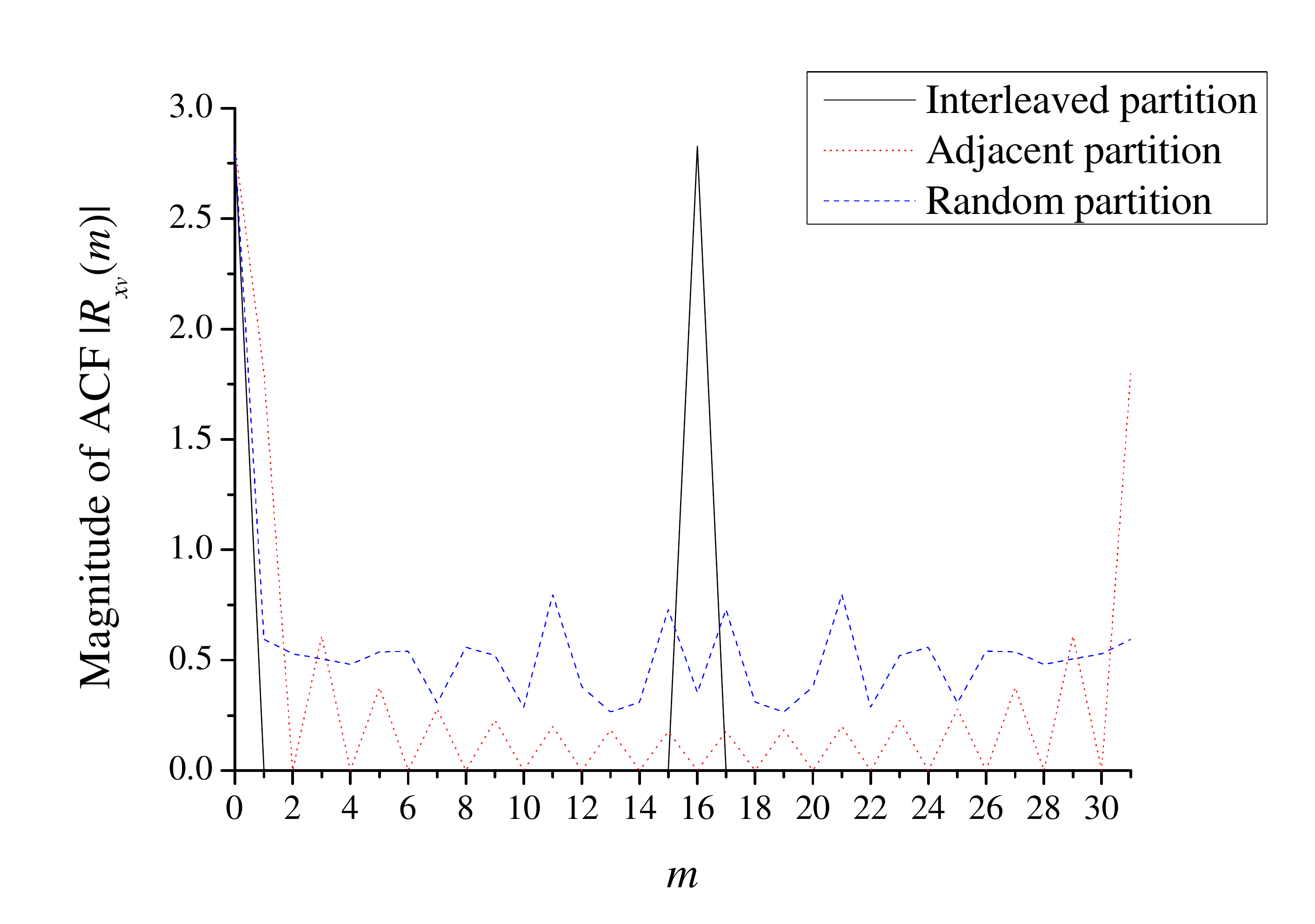}
\caption{Magnitude of ACFs for different partition cases.}
\label{fig:ACF}
\end{figure}

Fig. \ref{fig:ACF} shows an example of the magnitudes of ACFs corresponding to the following power spectrum when $N=32$ and $V=2$; $S_1 = \{ 1 0 1 0 \cdots 1 0 1 0 \}$ for an interleaved partition; $S_1 = \{ 1 1 \cdots 1 1 0 0 \cdots 0 0 \}$ for an adjacent partition; $S_1 = \{ 1 0 0 1 0 1 1 0 0 1 1 1 1 1 0 0 0 1 1 0 1 1 1 0 1 0 1 0 0 0 0 0 \}$ for a random partition, which is an one zero padded m-sequence with length 31; Clearly, $S_2$ is a complement of $S_1$ in each partition case, and the shapes of $|R_{x_v}(m)|$ for $v=1$ and $v=2$ are same.

\subsection{Desirable Shift Value Sets with Consideration of Correlation}
Now we investigate the desirable SV sets with consideration of correlation of the OFDM signal subsequence for three partition cases.
\subsubsection{For Random Partition}
In this case, the shape of the ACF is similar to a delta function. Therefore, the \emph{Criterion 1} can be valid criterion.

\subsubsection{For Interleaved Partition}
In this case, the shape of the ACF is the impulse train in (\ref{eq:intACF}). Then cyclic shift by $N/V$ cannot make the OFDM signal subsequence much different. Therefore, \emph{Criterion 1} has to be slightly modified as follows.

\textbf{\emph{Criterion 2}} : Suppose that we have $U$ SV sets; For every $(i,j)$ pair out of the $U$ SV sets ($i\neq j$), the pair should satisfy the condition that the relative distances $\tau_v^i - \tau_v^j \mod N/V$ are distinct from each other for all $v$'s.

\subsubsection{For Adjacent Partition}
In this case, the shape of the ACF in (\ref{eq:adACF}) is similar to a sinc function. Then cyclic shift by a small integer cannot make the OFDM signal subsequence much different. Instead, cyclic shift by an integer close to $N/2$ can make the OFDM signal subsequence much different because the magnitude of the ACF in (\ref{eq:adACF}) has a lower value as $m$ gets closer to $N/2$. For example, when $N=32$, as in Fig. \ref{fig:ACF}, the magnitude of the ACF has the lowest value when $m=16$. Therefore, the constraint that the relative distances have to be distinct from each other in \emph{Criterion 1} should be changed into a stronger constraint as follows.

\textbf{\emph{Criterion 3}} : Suppose that we have $U$ SV sets; For every $(i,j)$ pair out of the $U$ SV sets ($i\neq j$), the pair should satisfy the condition that the relative distances $\tau_v^i - \tau_v^j \mod N$ are distinct from each other for all $v$'s; Furthermore, the mutual differences of the $V$ relative distances ($\tau_1^i - \tau_1^j, \tau_2^i - \tau_2^j, \cdots, \tau_V^i - \tau_V^j   \mod N$) should be as close to $N/2$ as possible.

Unfortunately, it is very hard to describe the \emph{Criterion 3} more clearly, but it gives us an important insight to design the good SV sets for the adjacent partition case.

\section{Simulation Results}
\begin{figure}[h]
\centering
\includegraphics[width=\linewidth]{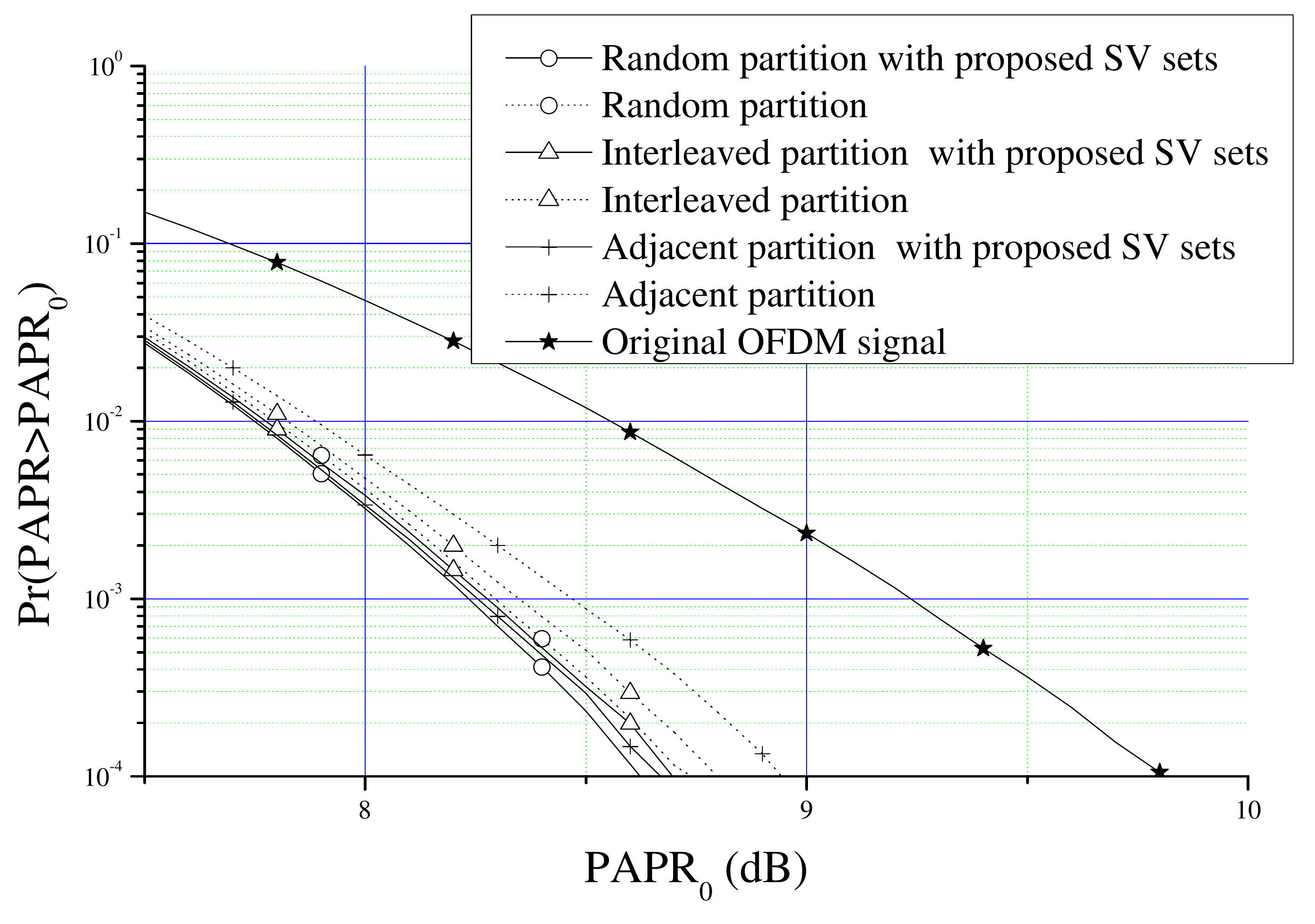}
\caption{Comparison of the PAPR reduction performance for three partition cases, which are random, interleaved, and adjacent partition cases.}
\label{fig:PAPR}
\end{figure}

To verify whether the above proposed criteria are valid, we construct the $U$ SV sets in two different ways. That is, the solid lines in Fig. \ref{fig:PAPR} show the PAPR reduction performance of the case when the $U$ SV sets satisfy the above criteria well. On the other hand, the dotted lines in Fig. \ref{fig:PAPR} show the PAPR reduction performance of the case that does not. In the simulations, we use $N=128$, $U=4$, and $V=4$ in common. The $16$-quadrature amplitude modulation (16-QAM) is used for all simulations.

\subsection{For Random Partition}
The SV sets $\overline{\tau}^1 = \{0,0,0,0\}$, $\overline{\tau}^2 = \{0,8,16,24\}$, $\overline{\tau}^3 = \{0,16,32,48\}$, and $\overline{\tau}^4 = \{0,24,48,72\}$ are used for the solid line, which satisfies \emph{Criterion 1}.
On the other hand, the SV sets $\overline{\tau}^1 = \{0,0,0,0\}$, $\overline{\tau}^2 = \{0,4,8,12\}$, $\overline{\tau}^3 = \{0,16,20,24\}$, and $\overline{\tau}^4 = \{0,28,32,36\}$ are used for the dotted line, which does not satisfy \emph{Criterion 1}. In Fig. \ref{fig:PAPR}, we can verify that the \emph{Criterion 1} for the random partition case is valid.

\subsection{For Interleaved Partition}
The SV sets $\overline{\tau}^1 = \{0,0,0,0\}$, $\overline{\tau}^2 = \{0,1,2,3\}$, $\overline{\tau}^3 = \{0,2,4,6\}$, and $\overline{\tau}^4 = \{0,3,6,9\}$ are used for the solid line, which satisfies \emph{Criterion 2}.
On the other hand, the SV sets $\overline{\tau}^1 = \{0,0,0,0\}$, $\overline{\tau}^2 = \{0,8,16,24\}$, $\overline{\tau}^3 = \{0,16,32,48\}$, and $\overline{\tau}^4 = \{0,24,48,72\}$ are used for the dotted line, which does not satisfy \emph{Criterion 2} (but still satisfies \emph{Criterion 1}). In Fig. \ref{fig:PAPR}, we can verify that the \emph{Criterion 2} for the interleaved partition case is valid.

\subsection{For Adjacent Partition}
The SV sets $\overline{\tau}^1 = \{0,0,0,0\}$, $\overline{\tau}^2 = \{0,44,73,95\}$, $\overline{\tau}^3 = \{0,9,35,84\}$, and $\overline{\tau}^4 = \{0,25,45,110\}$ are used for the solid line, which satisfies \emph{Criterion 3} well.
On the other hand, the SV sets $\overline{\tau}^1 = \{0,0,0,0\}$, $\overline{\tau}^2 = \{0,1,2,3\}$, $\overline{\tau}^3 = \{0,2,4,6\}$, and $\overline{\tau}^4 = \{0,3,6,9\}$ are used for the dotted line, which does not satisfy \emph{Criterion 3} well (but still satisfies \emph{Criterion 1} and \emph{Criterion 2}). In Fig. \ref{fig:PAPR}, we can verify that the \emph{Criterion 3} for the adjacent partition case is valid.

\section{Conclusion}
The CCS scheme is the very popular and promising PAPR reduction scheme, which is evolved from the PTS scheme. In this letter, the criteria to select good SV sets are proposed, which can guarantee the sub-optimal PAPR reduction performance of the CCS scheme. The criterion are proposed by considering the ACF of the OFDM signal subsequence for three different partition cases, random, interleaved, and adjacent partition cases. In the simulation results, the CCS scheme using the SV sets satisfying the proposed criteria shows better PAPR reduction performance than the case when the SV sets are not carefully designed.

\end{document}